\documentclass[runningheads]{llncs}
\usepackage[T1]{fontenc}
\usepackage[table,xcdraw]{xcolor}
\usepackage{graphicx}
\usepackage{subcaption}
\usepackage{booktabs}
\usepackage{multirow}
\usepackage{amssymb}
\usepackage{amsmath}
\usepackage{wrapfig,booktabs}

\usepackage[font=footnotesize,labelfont=bf]{caption}
\usepackage[english]{babel}
\usepackage{hyperref}
\usepackage[mode=buildnew]{standalone}
\usepackage{algorithm2e}
\begin{document}
\title{Comparing Ordering Strategies For Process Discovery Using Synthesis Rules}
%
%
\author{Tsung-Hao Huang \and
Wil M. P. van der Aalst}
\authorrunning{T. Huang and W. M. P. van der Aalst}
%
\institute{Process and Data Science (PADS), RWTH Aachen University, Aachen, Germany \\
\email{\{tsunghao.huang, wvdaalst\}@pads.rwth-aachen.de}\\
}
\maketitle              
\begin{abstract}
Process discovery aims to learn process models from observed behaviors, i.e., event logs, in the information systems.
The discovered models serve as the starting point for process mining techniques that are used to address performance and compliance problems.
Compared to the state-of-the-art Inductive Miner, the algorithm applying synthesis rules from the free-choice net theory discovers process models with more flexible (non-block) structures while ensuring the same desirable soundness and free-choiceness properties.
Moreover, recent development in this line of work shows that the discovered models have compatible quality.
Following the synthesis rules, the algorithm incrementally modifies an existing process model by adding the activities in the event log one at a time.
As the applications of rules are highly dependent on the existing model structure, the model quality and computation time are significantly influenced by the order of adding activities.
In this paper, we investigate the effect of different ordering strategies on the discovered models (w.r.t. fitness and precision) and the computation time using real-life event data. The results show that the proposed ordering strategy can improve the quality of the resulting process models while requiring less time compared to the ordering strategy solely based on the frequency of activities.

\keywords{Process discovery \and Synthesis rules \and Ordering strategy.}
\end{abstract}

\vspace{-1em}
\section{Introduction}\label{sec:intro}
\vspace{-0.5em}
Process mining, a discipline bridging the gap between process science and data science \cite{Aalst16PMbook}, offers techniques and tools to analyze event data, i.e., event logs, generated during the process execution.
The analysis generated by process mining techniques provides valuable data-driven insights for the stakeholders.

Process discovery is one of the three main research fields in process mining among conformance checking and process enhancement.
Process discovery techniques aim to learn end-to-end process models from the event data.
With the discovered models, knowledge workers can apply other process mining techniques to generate further insights for optimization.

While various algorithms have been proposed, only a few ensure desirable properties such as soundness and free-choiceness.
On the one hand, the soundness property guarantees that (1) it is always possible to finish the process (2) a process can be properly completed (3) no inexecutable transitions exist in the model~\cite{Aalst98worfklow}.
On the other hand, the free-choice property separates the choice and synchronization constructs of a process model (Petri net). 
Such property is desirable as it allows easy conversions from the discovered model to widely-used notations such as BPMN~\cite{Aalst21usingFreeChoice}.
Moreover, free-choice nets are supported by an abundance of analysis techniques developed from the theory~\cite{desel1995free}.

State-of-the-art techniques, such as the Inductive Miner (IM)~\cite{LeemansFA18scalableIM} family, discover process models guaranteed to be sound and free-choice. 
IM can provide such guarantees by exploiting its internal process representation - the \textit{process tree}.
However, such representation can also be a double-edged sword.
Due to the representational bias, the discovered models by IM are doomed to be block-structured, i.e., the model must compose of parts
that have a single entry and exit~\cite{LeemansFA18scalableIM}.
This implies that only a subset of sound free-choice workflow nets can be discovered by IM.

To provide a more flexible process representation while keeping the same guarantees, we proposed a novel discovery algorithm, the so-called Synthesis Miner in~\cite{Huang2022Pads_1st}. 
The Synthesis Miner utilizes the synthesis rules from the free-choice net theory~\cite{desel1995free}.
Activities in the event log are gradually added to a model under construction using predefined patterns.
Following the rules ensures that the discovered process models are always sound and free-choice.
Moreover, it is shown that the discovered models have compatible quality compared to the ones from Inductive Miner.
Nevertheless, the possible applications of synthesis rules are highly dependent on the existing model structure.
Different orders of adding activities can result in different models. 
Therefore, an open research question is the influence of the order in which the activities are added to an existing model on the final process model quality.
In this paper, we address the research question by comparing the ordering strategies for the Synthesis Miner and taking a deeper look into the impacts of the activity adding order to the model quality and computation time.
The experiment using four publicly available real-life event logs shows that advanced ordering strategies can significantly improve the model quality and the computation time.

The remainder of the paper is structured as follows. 
Related work is presented in Sect.\ref{sec:relatedWork}.
We introduce the necessary notations and concepts used throughout the paper in Sect.~\ref{sec:preliminaries}.
Then, the proposed ordering strategies are introduced in Sect.~\ref{sec:approach}.
The evaluation using publicly available real-life event logs is presented in~\ref{sec:evaluation}.
Finally, Sect.~\ref{sec:conclusion} concludes this paper.

\vspace{-1em}
\section{Related Work}\label{sec:relatedWork}
\vspace{-0.5em}
For a general introduction to process mining, we refer to \cite{Aalst16PMbook}.
Additionally, a review and benchmark of the recent development in process discovery can be found in~\cite{AugustoCDRMMMS19PDreviewbenchmark}.
In this paper, we focus on process discovery techniques that incrementally modify a model under construction to derive the final process.

Incremental process mining allows users to learn a process model from event logs by gradually integrating different traces into an existing model~\cite{SchusterZA20Incremental}.
As the ordering strategy has a significant impact on the model quality, a study~\cite{SchusterDZA22} is conducted to investigate the interplay.
Nevertheless, it is the trace that is added to the algorithm iteratively rather than the activity.
Therefore, it is less relevant to this paper.

Dixit et al.~\cite{DixitBA18ProDiGy} were among the first to use synthesis rules from free-choice net theory~\cite{desel1995free} to discover process models.
Inspired by~\cite{DixitBA18ProDiGy}, \cite{Huang2022Pads_1st} introduces the Synthesis Miner that automates the discovery by introducing predefined patterns and a search space pruning mechanism.
Both~\cite{DixitBA18ProDiGy} and~\cite{Huang2022Pads_1st} introduce a few ordering strategies for their approaches.
However, the choice of ordering is left to the user as an input parameter.
The impact of the ordering strategies on the model quality and computation time is not thoroughly investigated.
Furthermore, the interplay between the ordering strategies and the search space pruning has not been explained.
Last but not least, a comparison between different ordering strategies is needed.
In this paper, we aim to address the open research question and provide users with a rule of thumb.

\vspace{-1em}
\section{Preliminaries}\label{sec:preliminaries}
\vspace{-0.5em}
In this section, we introduce the necessary concepts and notations that are used throughout the paper.

For an arbitrary set $A$, we denote the set of all possible sequences as $A^{*}$ and the set of all multi-sets over $A$ as $\mathcal{B}(A)$.
Given $\sigma_1,\sigma_2 \in A^{*}$, $\sigma_1 \cdot \sigma_2$ denotes the concatenation of the two sequences.
Let $A$ be a set and $X \subseteq A$ be a subset of $A$.
For $\sigma \in A^{*}$ and $a \in A$, we define $\restriction_X \in A^{*} {\rightarrow} X^{*}$ as a projection function recursively with $\langle \rangle{\restriction_X} = \langle \rangle$, $(\langle a \rangle \cdot \sigma){\restriction_X} = \langle a \rangle \cdot \sigma{\restriction_X}$ if $a \in X$ and $(\langle a \rangle \cdot \sigma){\restriction_X} = \sigma{\restriction_X}$ if $a \notin X$.
For example, $\langle x,y,x \rangle{\restriction_{\{x,z\}}}=\langle x,x\rangle$.
The projection function can also be applied to a multi-set of sequences. 
For example, $[\langle x, y, x \rangle^{4}, \langle x, y \rangle^{2}, \langle y, x, z \rangle^{6}]{\restriction_{\{y,z\}}} = [\langle y\rangle^{6}, \langle y, z\rangle^{6}]$.
We denote $\mathcal{U}_{A}$ as the universe of activity labels.

\begin{definition}[Trace \& Log]
    A trace $\sigma \in \mathcal{U}_{A}^{*}$ is a sequence of activity labels.
    A log is a multi-set of traces, i.e., $L \in \mathcal{B}(\mathcal{U}_{A}^{*})$.
\end{definition}

\begin{definition}[Log Properties~\cite{Huang2022Pads_1st}]
    Let $L \in \mathcal{B}(\mathcal{U}_{A}^{*})$ and $a,b \in \mathcal{U}_{A}$ be two activity labels.
    We define the following log properties:
    \begin{itemize}
        \item $\#(a,L) = \Sigma_{\sigma \in L}|\{i \in \{1,2,...,|\sigma|\}| \sigma(i) = a\}|$ is the times $a$ occurred in $L$.
        \item $\#(a,b,L) = \Sigma_{\sigma \in L}|\{i \in \{1,2,...,|\sigma|-1\}| \sigma(i) = a\land \sigma(i+1) = b\}|$ is the number of direct successions from $a$ to $b$ in $L$.
        \item \begin{math}
            caus(a,b,L) = \begin{cases}
                              \frac{\#(a,b,L) - \#(b,a,L)}{\#(a,b,L) + \#(b,a,L) + 1}  & \text{if $a \neq b$}\\
                              \frac{\#(a,b,L)}{\#(a,b,L)+1} & \text{if $a = b$}
                        \end{cases}
        \end{math} is the strength of causal relation $(a,b)$.
        \item $A^{pre}_c(a,L) = \{a_{pre} \in \mathcal{U}_{A} | caus(a_{pre},a,L) \geq c\}$ is the set of $a$'s preceding activities, determined by threshold $c$.
        \item $A^{fol}_c(a,L) = \{a_{fol} \in \mathcal{U}_{A} | caus(a,a_{fol},L) \geq c\}$ is the set of $a$'s following activities, determined by threshold $c$.
    \end{itemize}
\end{definition}

    

\begin{definition}[Petri Net]
    Let $N=(P,T,F,l)$ be a Petri net, where P is the set of places, T is the set of transitions, $P \cap T = \emptyset$. $F\subseteq (P \times T) \cup (T \times P)$ is the set of arcs, and $l \in T \rightarrow \mathcal{U}_{A} \cup \{\tau\}$ is a labeling function that assigns activity labels to transitions. A transition $t\in T$ is invisible (or silent) if $l(t)=\tau$.
\end{definition}

\begin{definition}[Path \& Elementary Path]
    A path of a Petri net $N=(P,T,F)$ is a non-empty sequence of nodes $\rho = \langle x_1,x_2,...,x_n \rangle$ such that $(x_i,x_{i+1}) \in F$ for $1 \leq i < n$.
    $\rho$ is an elementary path if $x_i \neq x_j$ for $1\leq i<j\leq n$.
    For $X, X' \in P \cup T$, $\mathit{elemPaths}(X,X',N) \subseteq (P\cup T)^*$ is the set of all elementary paths from some $x\in X$ to some $x'\in X'$.
\end{definition}

\begin{definition}[Workflow Net (WF-net)~\cite{Aalst98worfklow}]\label{def:wf-net}
    Let $N=(P,T,F,l)$ be a Petri net. $W=(P,T,F,l,i,o,\top,\bot)$ is a WF-net iff 
    (1) it has a dedicated source place $i \in P$: $\bullet i=\emptyset$ and a dedicated sink place $o\in P$: $o\bullet = \emptyset$
    (2) $\top \in T$: $\bullet\top = \{i\}\land i\bullet = \{\top\}$ and $\bot \in T$: $\bot\bullet=\{o\}\land \bullet o = \{\bot\}$ 
    (3) every node $x$ is on some path from $i$ to $o$, i.e., $\forall_{x\in P\cup T} (i,x)\in F^* \land (x,o)\in F^*$, where $F^*$ is the reflexive transitive closure of $F$.
\end{definition}

\begin{definition}[Activity Order]
    Let $L \in \mathcal{B}(\mathcal{U}_{A}^{*})$ and $A=\bigcup_{\sigma\in L}\{a\in\sigma\}$.
    $\gamma\in A^{*}$ is an activity order for $L$ if $\{a\in\gamma\}=A$ and $|\gamma|=|A|$.
\end{definition}


\subsubsection{Synthesis Miner: Process Discovery Using Synthesis Rules}
In previous work~\cite{Huang2022Pads_1st}, we introduced the Synthesis Miner that guarantees to discover sound and free-choice workflow nets by applying the synthesis rules defined in~\cite{desel1995free} with an additional dual abstraction rule~\cite{Huang2022Pads_1st}.

\begin{figure}[h!]
    \centering
    \includegraphics[width=0.85\linewidth]{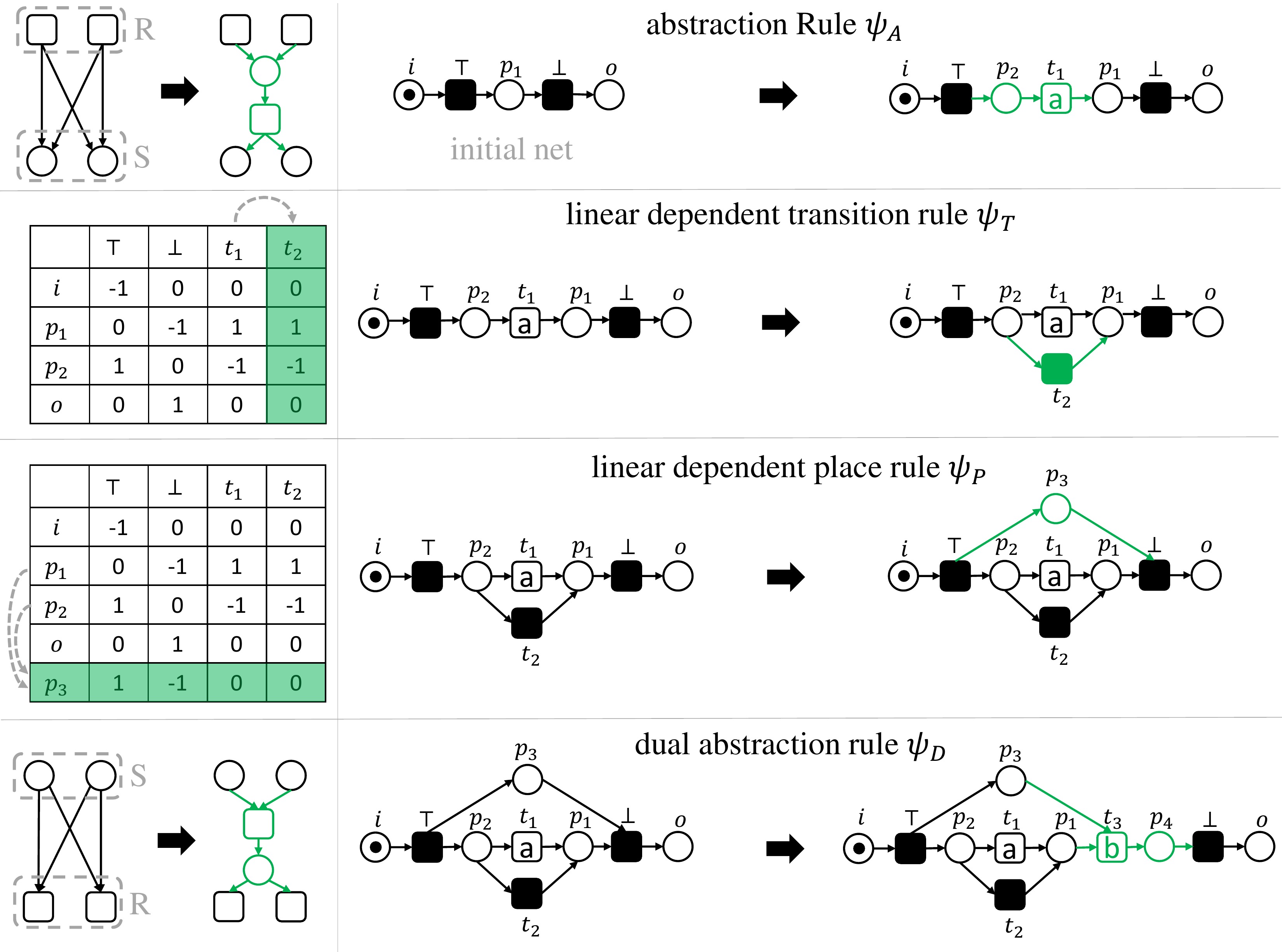}
    \caption{
    Some examples of the synthesis rules applications.
    $\psi_A$ allows to add $p_2$ and $t_1$ by $R=\{\top\}$ and $S=\{p_1\}$.
    $t_2$ is added by $\psi_T$ as it is linearly dependent on $t_1$.
    $p_3$ is added by $\psi_P$ as it is a linear combination of $p_1$ and $p_2$.
    $\psi_D$ allows to add $t_3$ and $p_4$ with $S=\{p_1,p_3\}$ and $R=\{\bot\}$.
    } \label{fig:rules_example}
    \vspace{-2em}
\end{figure}

Given a workflow net $W$, the abstraction rule ($\psi_A$) allows to add a place $p$ and a transition $t$ between a set of transitions $R\subseteq T$ and a set of places $S\subseteq P$ if they are fully connected, i.e., $(R\times S \subseteq F) \land (R\times S \neq \emptyset)$.
The linear transition/place rule ($\psi_T$/$\psi_P$) allows to add a transition $t$/place $p$ if it is linearly dependent on the other transitions/places in the corresponding incidence matrix.
The dual abstraction rule ($\psi_D$) can add a transition $t$ and a place $p$ between a set of places $S$ and a set of transitions $R$ if $(S\times R \subseteq F) \land (S\times R \neq \emptyset)$.
All four rules\footnote{For the formal definitions of the rules, we refer to~\cite{desel1995free,Huang2022Pads_1st}.} preserve sound and free-choice properties~\cite{desel1995free,Huang2022Pads_1st}.
Fig.~\ref{fig:rules_example} shows a few examples of rules applications.

Given a log $L$, the Synthesis Miner first determines an activity order $\gamma$.
Then, the iteration is initiated.
In iteration $i$ (where $1 \leq i\leq |\gamma|$), activity $\gamma(i)$ is added to an existing net\footnote{The existing net in the first iteration is initiated by the initial net, as shown in the example for the abstraction rule in Fig.\ref{fig:rules_example}.} from the $i-1$ iteration.
The procedure for every iteration is as follows: (1) use heuristics from the projected log $L_{i} = L{\restriction_{\{\gamma(1),\gamma(2),...\gamma(i)\}}}$ to find the most likely position for the to-be-added activity $\gamma(i)$ on the existing WF-net ($W_i$), 
(2) apply predefined patterns (derived from synthesis rules) to get the set of candidate nets, and 
(3) select the best net (w.r.t. fitness and precision) from the set of candidates for the next iteration.

As step (1) is directly affected by the ordering strategy, we formally define\footnote{As the formal definitions of steps (2) and (3) are out of scope, we refer to~\cite{Huang2022Pads_1st}.} how the search space is limited to only a subset of the nodes on a workflow net using log heuristics.

\begin{definition}[Reduced Search Space]\label{def:reduced-search-space}
    Let $a\in \mathcal{U}_{A}^{*}$ be an activity, $L {\in} \mathcal{B}(\mathcal{U}_{A}^{*})$ be a log, $W=(P,T,F,l,i,o,\top,\bot)$ be a WF-net, and $0\leq c\leq 1$.
    $T^{pre}$ is the set of transitions labeled by the preceding activities of $a$ in log $L$.
    $T^{pre}=\{t\in T | l(t)\in A^{pre}_c(a,L)\}$ if $A^{pre}_c(a,L) \neq \emptyset$, otherwise $T^{pre}=\{\top\}$.
    $T^{fol}$ is the set of transitions labeled by the following activities of $a$ in log $L$.$T^{fol}=\{t\in T | l(t)\in A^{fol}_c(a,L)\}$ if $A^{fol}_c(a,L) \neq \emptyset$, otherwise $T^{fol}=\{\bot\}$.
    The reduced search space is $\mathit{reduce}(a,L,W,c)=\{x\in \rho |\rho\in\mathit{elemPath}(T^{pre},T^{fol},W)\}$.
\end{definition}
\begin{figure}[h!]
    \begin{subfigure}[b]{0.49\linewidth}
        \includegraphics[width=\linewidth]{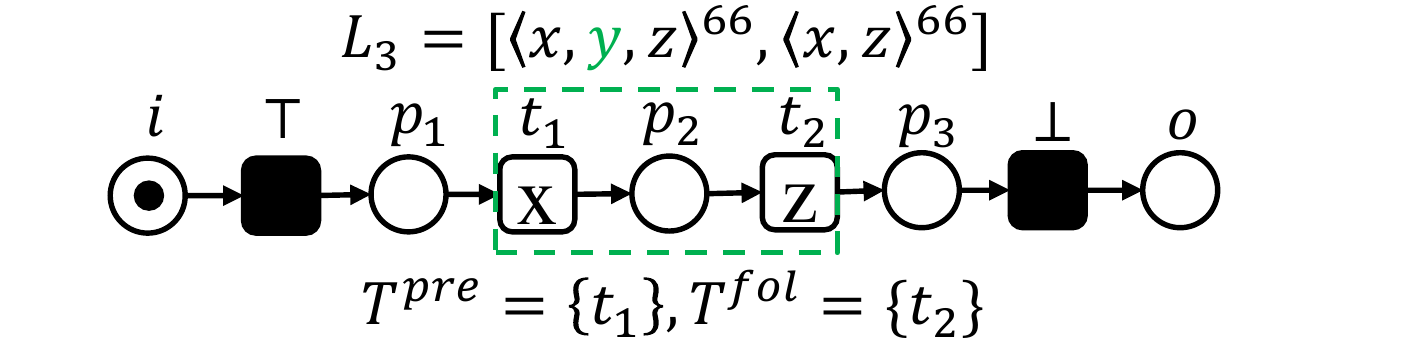}
        \caption{$W_2$, the existing net from the last iteration}\label{subfig:prune_example_1}
    \end{subfigure}
    \begin{subfigure}[b]{0.49\linewidth}
        \includegraphics[width=\linewidth]{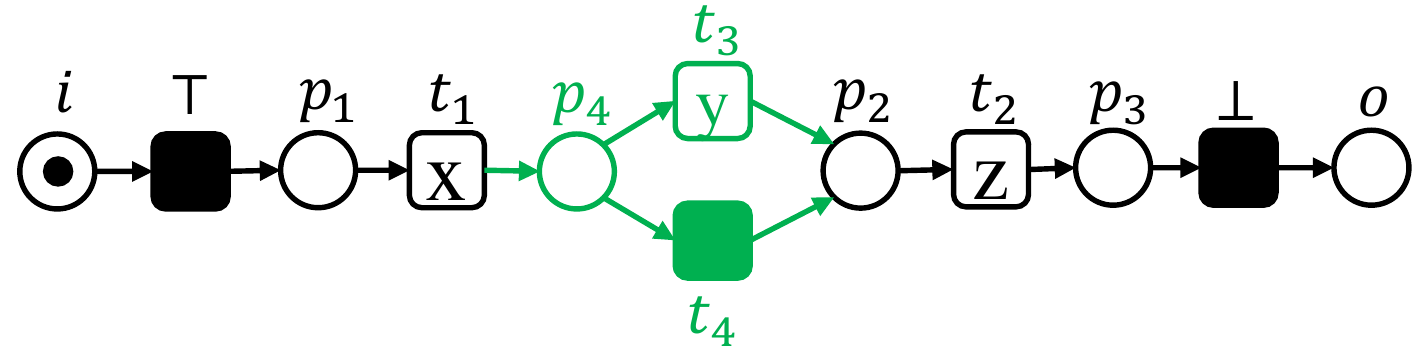}
        \caption{$W_3$, the net after adding $y$}\label{subfig:prune_example_2}
    \end{subfigure}
    \caption{An example showing how the search space is reduced. Consider the log $L_3=[\langle x,y,z\rangle^{66},\langle x,z\rangle^{66}]$. 
    $y$ is the activity which we want to add to the net $W_2$.
    Using $c=0.9$, we get $T^{pre}=\{t_1\}$ and $T^{fol}=\{t_2\}$. Therefore, the function $reduce$ would return the set of nodes between $t_1$ and $t_2$, which means $V_3 = \{t_1,p_2,t_2\}$ as highlighted by the green dashed line in (a). 
    The application of synthesis rules would then only consider these three nodes.
    Finally, the best net is selected as $W_3$ from the candidates and is visualized in (b).
    }
    \label{fig:prune_example}
    \vspace{-1em}
\end{figure}
The function $\mathit{reduce}$ first finds the preceding and following activities and the corresponding sets of labeled transitions for the to-be-added activity $\gamma(i)$. 
Then, it returns the set of nodes, denoted as $V_i$, that are on the path between the preceding and following transitions.
$V_i$ is used to confine the application of synthesis rules.
To be more precise, the set of transitions $R$ and the set of places $S$ used as the preconditions for applying rules $\psi_A$ and $\psi_D$ need to be a subset of $V_i$, i.e., $S\subseteq V \land R\subseteq V$.
As for rule $\psi_T$/$\psi_P$, the new transition/place ($t'$/$p'$) cannot have arcs connected to any node other than $V_i$.
This step helps us to limit the search space to the most likely nodes on a workflow net to add activity $\gamma(i)$.
Fig.~\ref{fig:prune_example} shows an example for reducing the search space.

\vspace{-0.5em}
\section{Ordering Strategies}\label{sec:approach}
\vspace{-0.5em}
In this section, we introduce different ordering strategies.
To illustrate the ordering strategy, consider the following log
$L_s=[\langle b,c,d,e,f,g\rangle,
    \langle b,e,c,d,f,g\rangle,
    \langle b,e,c,\\f,g,d\rangle,
    \langle b,e,c,f,d,g\rangle,
    \langle b,c,e,d,f,g\rangle,
    \langle b,c,e,f,g,d\rangle,
    \langle b,c,e,f,d,g\rangle,
    \langle e,b,c,d,f,g\rangle,\\
    \langle e,b,c,f,g,d\rangle,
    \langle e,b,c,f,d,g\rangle]$.
\begin{wrapfigure}[6]{r}{.5\textwidth}
    \centering
    \vspace{-3em}
    \includegraphics[width=\linewidth]{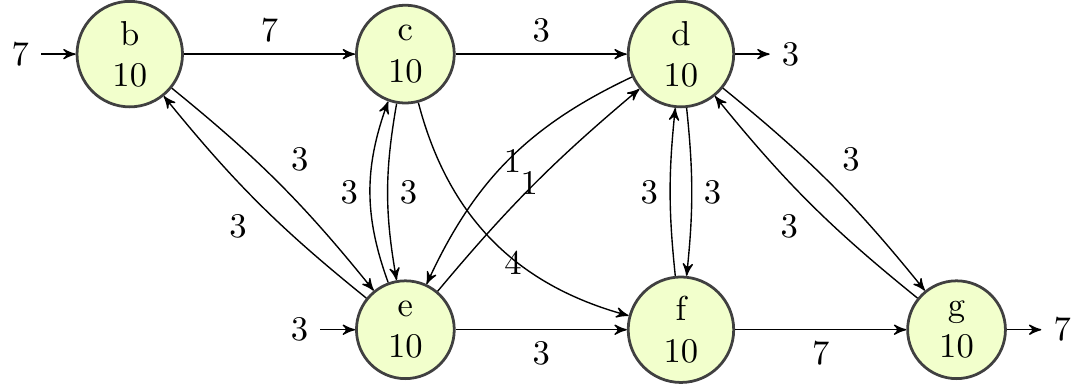}
    \caption{The DFG for log $L_s$. } \label{fig:DFG}
\end{wrapfigure}
The corresponding directly follows graph (DFG) is shown in Fig.~\ref{fig:DFG}.


The first ordering strategy is frequency-based and it is relatively straightforward. 
The activities are simply ordered by their frequency in the log.

\begin{definition}[Frequency-Based Ordering]
    Let $L{\in}\mathcal{B}(\mathcal{U}_{A}^{*})$.
    Frequency-based ordering function is $\mathit{order_{freq}}(L) = \gamma$ such that $\gamma$ is an activity order and $\forall_{1\leq i<j\leq|\gamma|} \#(\gamma(i),L) \geq \#(\gamma(j),L)$.
\end{definition}
If activities have the same frequency, we order them alphabetically.
Using the example log $L_s$ for illustration, the order would be $\mathit{order_{freq}}(L_s)=\langle b,c,d,e,f,g\rangle$.

The other ordering strategies are more involved as they consider not only the frequency of activities but also the connections between them.
Before introducing the other ordering strategies, we first define a helper function that ranks the directly-follow activities based on the strength of connections.


\begin{definition}[Directly-Follow Activities Sorting]\label{def:directly-follow-activities-sorting}
    Let $L {\in} \mathcal{B}(\mathcal{U}_{A}^{*})$ and $a {\in} \mathcal{U}_{A}$. $A {=} \{b\in\mathcal{U}_{A}|\#(a,b,L){>}0\}$ is the set of activities directly-follow $a$ in $L$ at least once and $\sigma\in A^*$.    Directly-follow activities sorting is $\mathit{sortDFA}(a,L) = \sigma$ such that $\{b\in\sigma\}=A$ and $|\sigma|=|A|$ and $\forall_{1\leq i<j\leq |\sigma|} \,  \#(a,\sigma(i),L) \geq \#(a,\sigma(j),L)$.
\end{definition}
For example, $\mathit{sortDFA}(b,L_s)=\langle c,e\rangle$. 
This is because activities $c$ and $e$ have incoming arcs from $b$ and the strength $\#(b,c,L_s) \geq \#(b,e,L_s)$.
With the function for sorting directly-follow activities defined, we are now ready to define the Breadth-First-Search-Based ordering strategy in Algo.~\ref{alg:BFS-ordering}.

\RestyleAlgo{ruled}

\newcommand\mycommfont[1]{\scriptsize\ttfamily\textcolor{teal}{#1}}
\SetCommentSty{mycommfont}
\SetKwInOut{Input}{Input}\SetKwInOut{Output}{Output}
\SetKwFor{While}{while}{:}{}%
\SetKwFor{If}{if}{:}{}%
\SetKwFor{Else}{else}{:}{}%

\begin{algorithm}[hbt!]
    \caption{Breadth-First-Search-Based Ordering, $\mathit{order_{BFS}}$}\label{alg:BFS-ordering}
    \Input{A log $L \in \mathcal{B}(\mathcal{U}_{A}^{*})$}
    \Output{An activity order $\gamma$ for $L$}
    $A \gets \bigcup_{\sigma\in L}\{a\in\sigma\}$ \tcp*{the set of activities in $L$}
    $A^{s}\gets \{\sigma(1) \, |\, \sigma \in L \land \sigma \neq \langle\rangle \}$ \tcp*{the set of start activities in $L$}
    $\sigma \gets \mathit{order_{freq}}(L){\restriction_{A^{s}}}$ \tcp*{the sequence of start activities ordered by frequency}
    $i \gets 1$\;
    \While{$|\sigma| \neq |A|$}{
        $A' \gets A\setminus\{a \in \sigma \}$ \tcp*{the set of activities that are not in $\sigma$}
        $\sigma' \gets \mathit{sortDFA}(\sigma(i), L){\restriction_{A'}}$ \tcp*{sort $\sigma(i)$'s following activities \& project on $A'$}
        $\sigma \gets \sigma \cdot \sigma'$ \tcp*{update $\sigma$}
        $i \gets i + 1$\;
    }
    $\gamma \gets \sigma$\;
    \Return{$\gamma$}\;
\end{algorithm}

BFS-based ordering strategy starts by building a sequence of start activities in a log and iteratively append the sequence of directly-follow activities using the function in Def.~\ref{def:directly-follow-activities-sorting}.
Applying the function to the example log $L_{s}$, we get $\mathit{order_{BFS}(L_s)}=\langle b,e \rangle\cdot\langle c\rangle\cdot \langle f,d\rangle\cdot\langle\rangle\cdot\langle g\rangle=\langle b,e,c,f,d,g\rangle$.
$\sigma$ is initiated with $\langle b,e\rangle$.
Then, in iteration $i$, $\sigma$ is appended by the sequence of $\sigma(i)$'s directly-follow activities sorted by $\mathit{sortDFA}(\sigma(i), L_s)$ with the set of activities already in $\sigma$ filtered out.
The loop continues until $\sigma$ includes every activity in the log.
As its name suggests, the ordering prioritizes the exploration of the directly-follow activities.

Next, we introduce another ordering strategy in Algo.~\ref{alg:DFS-ordering} that is Depth-First-Search-based.
\begin{algorithm}[h!]
    \caption{Depth-First-Search-Based Ordering $\mathit{order_{DFS}}$}\label{alg:DFS-ordering}
    \Input{A log $L \in \mathcal{B}(\mathcal{U}_{A}^{*})$}
    \Output{An activity order $\gamma$ for $L$}
    $A \gets \bigcup_{\sigma\in L}\{a\in\sigma\}$ \tcp*{the set of activities in $L$}
    $A^{s}\gets \{\sigma(1) \, |\, \sigma \in L \land |\sigma| \neq 0 \}$ \tcp*{the set of start activities in $L$}
    $\sigma^s \gets \mathit{order_{freq}}(L){\restriction_{A^{s}}}$ \tcp*{the sequence of start activities ordered by frequency}
    $\sigma \gets \langle \sigma^s(1) \rangle$ \tcp*{initiate the sequence with the most frequent start activity}
    $\sigma^s \gets \sigma^s{\restriction_{\{A^s\setminus\{\sigma^s(1)\}\}}}$ \tcp*{update $\sigma^s$ to be the stack}
    \While{$|\sigma| \neq |A|$}{
        $A' \gets A\setminus\{a\in\sigma\}$ \tcp*{set of activities that are not in $\sigma$}
        $\sigma^f \gets \mathit{sortDFA}(\sigma(|\sigma|), L){\restriction_{A'}}$ \tcp*{sort $\sigma(|\sigma|)$'s following activities}
        \If{$|\sigma^f|=0$}{
            $\sigma \gets \sigma\cdot\langle\sigma^s(1)\rangle$ \tcp*{append the 1st element from the stack $\sigma^s$ to $\sigma$ }
        }
        \Else{}{
            $\sigma \gets \sigma\cdot\langle\sigma^f(1)\rangle$ \tcp*{append the 1st element from $\sigma^f$ to $\sigma$ }
        }
        $\sigma^s \gets (\sigma^f{\restriction_{A\setminus\{a\in\sigma \lor a\in\sigma^s\}}})\cdot(\sigma^s{\restriction_{A\setminus\{a \in \sigma\}}})$ \tcp*{update the stack $\sigma^s$}
    }
    $\gamma \gets \sigma$\;
    \Return{$\gamma$}\;
\end{algorithm}
While also considering the connection between the activities as BFS-based ordering strategy, DFS-based ordering prioritizes depth over breadth. 
That is, the directly-follow activities are not explored thoroughly until activities with higher depth have been explored.
Applying DFS-based ordering to log $L_s$, we get $\mathit{order_{DFS}}(L_s)=\langle b,c,f,g,d,e\rangle$.

Note that although we define the BFS- and DFS-based ordering strategies to start from the start activities, one can also initiate the exploration from another direction, i.e., from the end activities and subsequently explore the directly-precede activities for ordering.
Using $L_s$ as an example, if starting from the set of end activities, we would get $\langle g,f,c,b,e,d\rangle$ with DFS-based ordering on log $L_s$ and $\langle g,d,f,c,e,b\rangle$ with BFS-based ordering.

\begin{figure}[h!]
    \centering
    \includegraphics[width=\linewidth]{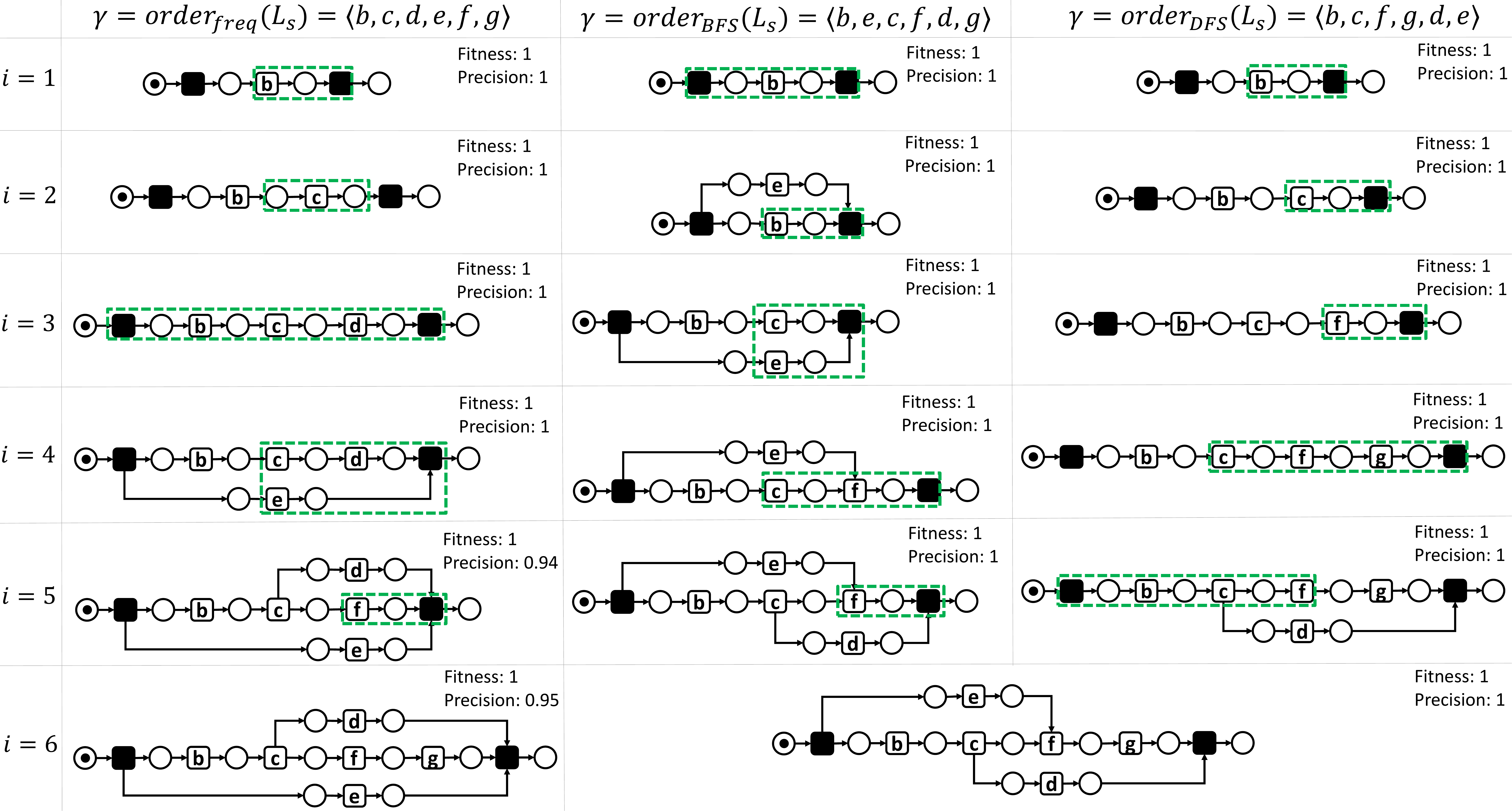}
    \caption{A comparison of different ordering strategies for log $L_s$. Each column represents an ordering strategy and each row corresponds to the intermediate workflow net in iteration $i$ after adding $\gamma(i)$. The green dashed lines highlight the nodes representing the reduced search space. 
    The metrics fitness and precision are measured using the corresponding projected log $L_{i} = L{\restriction_{\{\gamma(1),\gamma(2),...\gamma(i)\}}}$.
    Note that the final model discovered by the BFS- and DFS-based ordering strategies are the same in this example.} \label{fig:ordering_comparison}
    \vspace{-1.5em}
\end{figure}

To explain how the progression of the process discovery influenced by the different ordering strategies, Fig.~\ref{fig:ordering_comparison} shows all the intermediate nets when applying Synthesis Miner to log $L_s$ using the three different ordering strategies.
DFS-based ordering tends to build the process from start to end at the beginning before adding the activities in the parallel/choice branches.
On the contrary, BFS-based ordering prioritizes the construction of local control flows.
For example, the difference is observable from iteration 1 to 2.
While all the ordering strategies produce the same net in iteration 1, BFS-based ordering suggests to add the concurrent activity $e$ for $b$ in iteration 2 and DFS-based ordering adds the directly-follow activity $c$ of $b$ first.
The frequency ordering doesn't seem to have clear patterns for the discovery.

We expect that the choice of ordering can significantly influence the computation time of discovery.
The main difference stems from the time required to check the feasibility of the linear dependency rules.
As the WF-net grows, it becomes more expensive (w.r.t. time) to check if a candidate place/transition is linear dependent.
Thus, it is preferable to limit the search space as small as possible, especially in the later iterations.
Recall that the reduced search space (Def.~\ref{def:reduced-search-space}) is a set of nodes confining the application of synthesis rules.
The green dashed lines in Fig.~\ref{fig:ordering_comparison} highlight the reduced search space $V_i$ in iteration $i$.
As shown in Fig.~\ref{fig:ordering_comparison}, generally, BFS-based ordering can keep the search space smaller than the other strategies because it prioritizes the connected activities.
In contrast, the search space of DFS-based ordering is more likely to be large in the later iterations.
As the parallel/alternative activities are added later, the preceding and following activities of the to-be-added activity $\gamma(i)$ is highly likely to be spread across the existing net.
Together with the effect of search space reduction, it results in a relatively large search space, which indicates more nodes to be considered.
Examples can be seen in iterations 4 and 5 for the DFS-based ordering in Fig.~\ref{fig:ordering_comparison}.

Although it is assumed that BFS-based ordering would have relatively lower computation time, search space reduction might introduce trade-offs between the optimal solution and time.
In the following section, we aim to investigate the impact of the ordering strategy on both model quality and the time to discover the process model in the experiment.

\vspace{-0.5em}
\section{Evaluation}\label{sec:evaluation}
\vspace{-0.5em}
In this section, we present the experiment used to evaluate the ordering strategies including the setup and a discussion of the result\footnote{ \url{https://github.com/tsunghao-huang/synthesisRulesMiner}}.

\vspace{-0.5em}
\subsection{Experimental Setup}
\vspace{-0.5em}
For the experiment, we use four publicly available real-life event logs~\cite{vanDongen2017BPI2017,deLeoni2015RTFM,Mannhardt2017Hospital,Polato2017Helpdesk}.
The logs are filtered to focus on the mainstream behaviors (at least 95\% of the traces) where the most frequent trace variants are used.
For the BPI2017 log~\cite{vanDongen2017BPI2017}, we split it into three logs using the activity prefix (A, W, O). 
This results in six logs in total.

For every event log, we apply different ordering strategies for the Synthesis Miner~\cite{Huang2022Pads_1st} with default values for the other parameters.
For the BFS- and DFS-based ordering strategies, we apply the ordering from both directions (start and end activities). 
Therefore, we evaluate five ordering strategies.
To measure the effect of ordering strategies on search space pruning, we keep track of the ratio of reduced search space.
This is evaluated by $\frac{|V_i|}{|P_i\cup T_i| - 2}$, where $V_i$ is the set of reduced nodes, $P_i$ and $T_i$ are the set of places and transitions in the existing WF-net $W_i$. 
The $-2$ in the denominator is there to exclude the two places (source and sink) that can never be connected by new nodes by Def.~\ref{def:wf-net}.
Using Fig.~\ref{fig:ordering_comparison} as an example, the value of $\frac{|V_3|}{|P_3\cup T_3| - 2}$ for the frequency ordering strategy would be $\frac{9}{11-2}=1$ in iteration 3.
This indicates that all the possible nodes are considered for the application of synthesis rules to add the next activity.
Furthermore, we evaluate the final model in terms of fitness, precision, and F1 score (the harmonic mean of fitness and precision).

\subsection{Results and Discussion}
\subsubsection{Search Space Reduction and Computation Time}
\begin{figure}[h!]
    \vspace{-2em}
    \centering
        \begin{subfigure}[b]{.49\linewidth}
            \includegraphics[width=\linewidth]{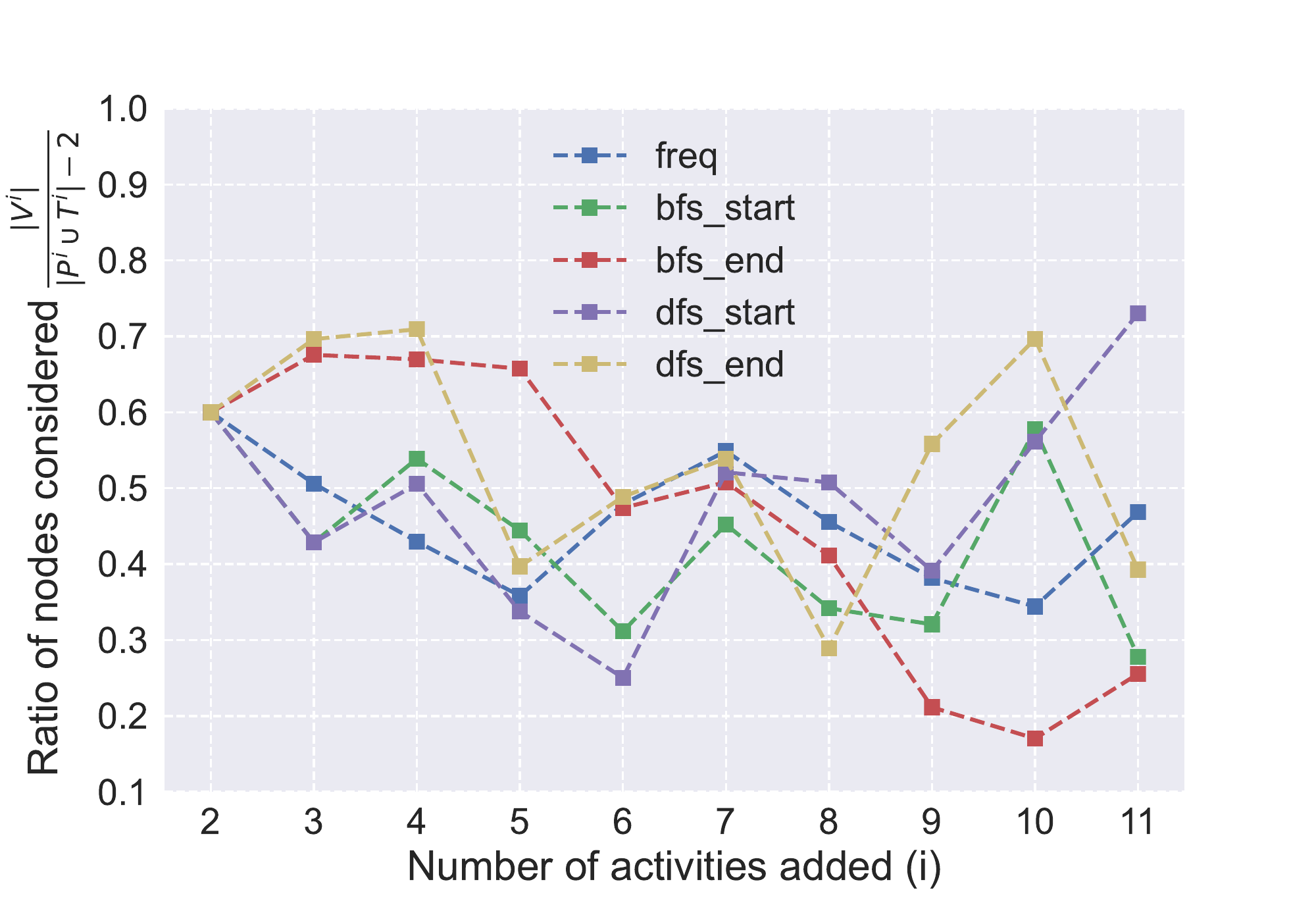}
            \caption{Average ratio of reduced search space}\label{fig:nodes_ratio_avg}
        \end{subfigure}
        \begin{subfigure}[b]{.49\linewidth}
            \includegraphics[width=\linewidth]{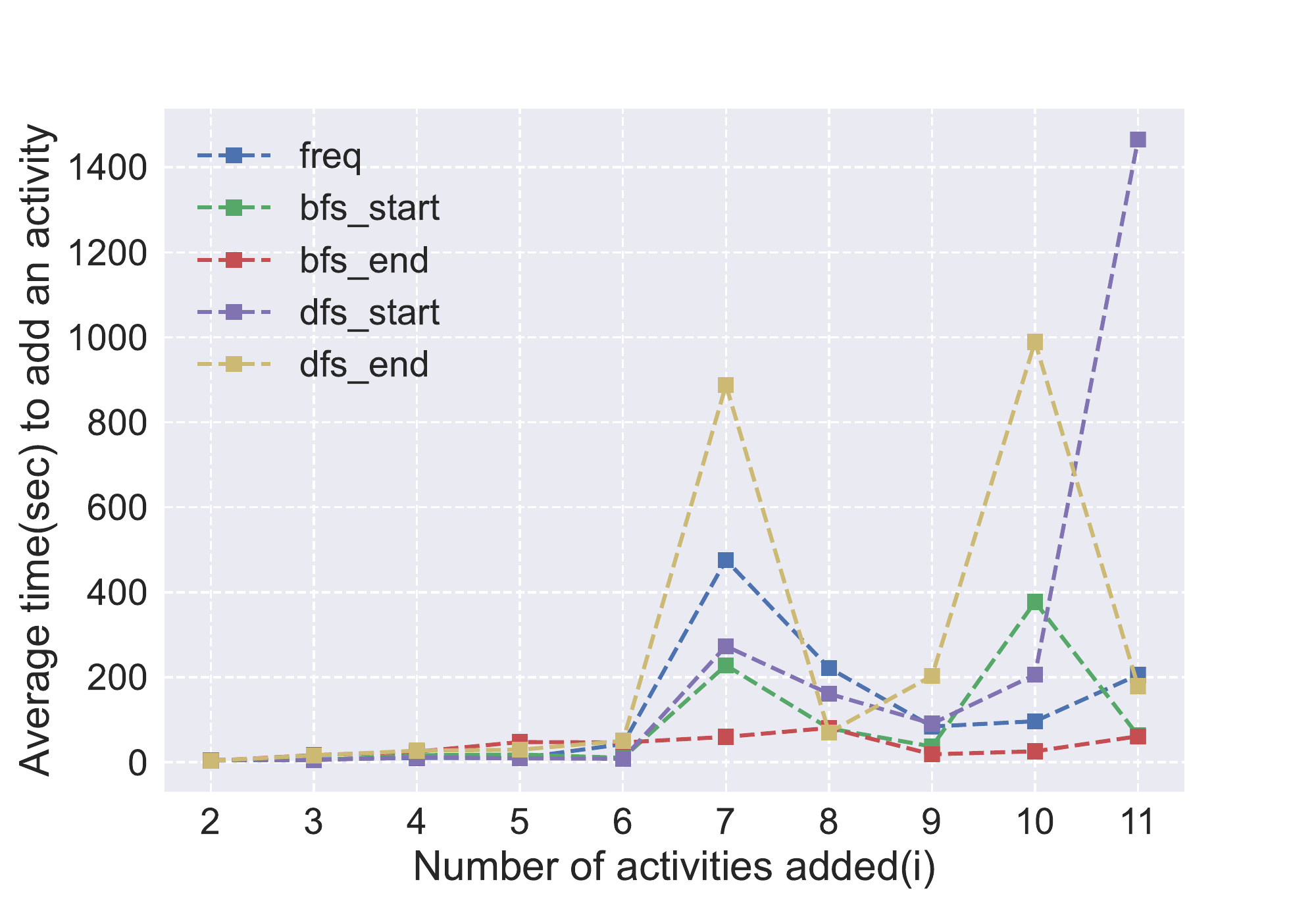}
            \caption{Average time to add an activity}\label{fig:avg_time_to_add_activity}
        \end{subfigure}
    \caption{Comparisons of ordering strategies on the effects of search space reduction as well as the computation time for each step.
    Note that it is preferable to have a lower value for $\frac{|V_i|}{|P_i\cup T_i| - 2}$.}
    \label{fig:result_nodes_ratio}
    \vspace{-2em}
\end{figure}

Fig.~\ref{fig:result_nodes_ratio} shows the result of the comparison among the five ordering strategies regarding their effects on the search space reduction.
The value in the y-axis $\frac{|V_i|}{|P_i\cup T_i| - 2}$ is the average across six event logs.
As indicated, the metric keeps track of the reduced search space ratio for adding the next activity, which indicates the number of possible synthesis rule applications.
In general, we can observe from the figure that the ordering strategies behaved as expected. 
As shown in Fig.~\ref{fig:nodes_ratio_avg}, in the later stage of the discovery ($i\geq 8$), the BFS-ordering strategies (bfs\_start, bfs\_end) keep the ratio of reduced search space at a low level while the value for frequency and DFS-based ordering strategies show that they are more likely to include a large portion of the nodes in the search space.

Fig.~\ref{fig:avg_time_to_add_activity} shows the average time to add an activity to the existing WF-net for each step of six logs.
Comparing the two figures, one can see the effect of search space reduction on the computation time.
As shown in Fig.~\ref{fig:avg_time_to_add_activity}, the bfs\_end strategy keeps the average computation time for each step at a fairly low level.
This is also the case for the bfs\_start strategy despite the two peaks when adding the 7th and 10th activity.
The two peaks in the 7th and 10th steps are especially severe for the dfs\_end strategy. 
Both took more than 10 minutes to add a single activity to the existing model.
Also, the longest duration to add an activity also happens in the 11th step of the dfs\_start strategy.

In short, due to its interplay with the search space reduction, the BFS-based ordering strategies have significant advantage in terms of computation time.

\begin{table}[t]
\centering
\caption{Quality of the models discovered by different ordering strategies.}
\label{tab:model_quality}
\resizebox{0.60\textwidth}{!}{%
\begin{tabular}{c|c|c|c|c|c}
\hline
Log &
  Ordering Strategy {\color[HTML]{9B9B9B} \& IMf} &
  Fitness &
  Precision &
  F1 &
  time(sec) \\ \hline \hline
 &
  frequency &
  0.971 &
  0.947 &
  0.958 &
  685 \\
 &
  BFS\_start &
  0.973 &
  1.000 &
  0.986 &
  893 \\
 &
  BFS\_end &
  0.990 &
  0.935 &
  0.961 &
  \textbf{334} \\
 &
  DFS\_start &
  0.963 &
  0.868 &
  0.913 &
  1850 \\
 &
  DFS\_end &
  0.999 &
  0.986 &
  \textbf{0.993} &
  1248 \\
\multirow{-6}{*}{BPI2017A} &
  {\color[HTML]{9B9B9B} IMf(0.2)} &
  {\color[HTML]{9B9B9B} 0.999} &
  {\color[HTML]{9B9B9B} 0.936} &
  {\color[HTML]{9B9B9B} 0.967} &
  {\color[HTML]{9B9B9B} 10} \\ \hline
 &
  frequency &
  0.993 &
  0.962 &
  0.978 &
  537 \\
 &
  BFS\_start &
  0.985 &
  0.963 &
  0.974 &
  \textbf{165} \\
 &
  BFS\_end &
  0.989 &
  1.000 &
  0.995 &
  231 \\
 &
  DFS\_start &
  0.996 &
  1.000 &
  \textbf{0.998} &
  498 \\
 &
  DFS\_end &
  0.993 &
  0.962 &
  0.978 &
  360 \\
\multirow{-6}{*}{BPI2017O} &
  {\color[HTML]{9B9B9B} IMf(0.2)} &
  {\color[HTML]{9B9B9B} 0.997} &
  {\color[HTML]{9B9B9B} 0.907} &
  {\color[HTML]{9B9B9B} 0.950} &
  {\color[HTML]{9B9B9B} 7} \\ \hline
 &
  frequency &
  0.993 &
  0.726 &
  0.838 &
  3617 \\
 &
  BFS\_start &
  0.974 &
  0.864 &
  0.914 &
  1626 \\
 &
  BFS\_end &
  0.993 &
  0.888 &
  0.936 &
  \textbf{579} \\
 &
  DFS\_start &
  0.974 &
  0.864 &
  0.914 &
  1732 \\
 &
  DFS\_end &
  0.993 &
  0.901 &
  \textbf{0.944} &
  5397 \\
\multirow{-6}{*}{BPI2017W} &
  {\color[HTML]{9B9B9B} IMf(0.2)} &
  {\color[HTML]{9B9B9B} 0.923} &
  {\color[HTML]{9B9B9B} 0.897} &
  {\color[HTML]{9B9B9B} 0.910} &
  {\color[HTML]{9B9B9B} 14} \\ \hline
 &
  frequency &
  0.974 &
  0.984 &
  0.978 &
  51 \\
 &
  BFS\_start &
  0.974 &
  0.984 &
  0.978 &
  52 \\
 &
  BFS\_end &
  0.983 &
  0.976 &
  \textbf{0.979} &
  \textbf{43} \\
 &
  DFS\_start &
  0.974 &
  0.984 &
  0.978 &
  49 \\
 &
  DFS\_end &
  0.989 &
  0.963 &
  0.976 &
  64 \\
\multirow{-6}{*}{helpdesk} &
  {\color[HTML]{9B9B9B} IMf(0.2)} &
  {\color[HTML]{9B9B9B} 0.967} &
  {\color[HTML]{9B9B9B} 0.950} &
  {\color[HTML]{9B9B9B} 0.958} &
  {\color[HTML]{9B9B9B} 1} \\ \hline
 &
  frequency &
  0.945 &
  0.810 &
  0.879 &
  509 \\
 &
  BFS\_start &
  0.931 &
  0.922 &
  0.936 &
  \textbf{314} \\
 &
  BFS\_end &
  0.988 &
  0.935 &
  \textbf{0.961} &
  383 \\
 &
  DFS\_start &
  0.931 &
  0.970 &
  0.961 &
  2154 \\
 &
  DFS\_end &
  0.943 &
  0.883 &
  0.920 &
  2359 \\
\multirow{-6}{*}{\begin{tabular}[c]{@{}c@{}}hospital \\ billing\end{tabular}} &
  {\color[HTML]{9B9B9B} IMf(0.2)} &
  {\color[HTML]{9B9B9B} 0.982} &
  {\color[HTML]{9B9B9B} 0.906} &
  {\color[HTML]{9B9B9B} 0.943} &
  {\color[HTML]{9B9B9B} 45} \\ \hline
 &
  frequency &
  0.967 &
  0.930 &
  0.945 &
  274 \\
 &
  BFS\_start &
  0.967 &
  0.930 &
  0.945 &
  \textbf{202} \\
 &
  BFS\_end &
  0.972 &
  0.720 &
  0.825 &
  388 \\
 &
  DFS\_start &
  0.991 &
  0.933 &
  \textbf{0.960} &
  366 \\
 &
  DFS\_end &
  0.942 &
  0.858 &
  0.903 &
  443 \\
\multirow{-6}{*}{traffic} &
  {\color[HTML]{9B9B9B} IMf(0.4)} &
  {\color[HTML]{9B9B9B} 0.904} &
  {\color[HTML]{9B9B9B} 0.720} &
  {\color[HTML]{9B9B9B} 0.801} &
  {\color[HTML]{9B9B9B} 28} \\ \hline
\end{tabular}%
}
\vspace{-1.5em}
\end{table}

\vspace{-0.5em}
\subsubsection{Model Quality}
Table~\ref{tab:model_quality}\footnote{To provide a reference to the state of the art, we also present the results from IMf (marked by gray color).
The best model generated by IMf (w.r.t. F1 score) is selected from a set of nets using five different values ($[0.1,0.2,0.3,0.4,0.5]$) for the filter.
} shows the result of the model quality using the five different ordering strategies.
As expected, we observe that the BFS-based ordering strategies have the lowest computation time in all six event logs.
This corresponds to the findings in the previous section.
Moreover, despite the search space being considerably reduced, the models discovered using BFS-ordering strategies have the highest F1 score in two out of the six logs.

As for the DFS-based ordering strategies, they have an apparent disadvantage for computation time but get the highest F1 score in the other four event logs.
The result matches our assumption as search space reduction introduces a trade-off between the optimal solution and time.
Lastly, the frequency ordering strategy has no significant advantage in model quality and computation time.
The results show that the ordering strategies that take the connections between activities into consideration can improve the Synthesis Miner than the frequency-based ordering strategy. 

\vspace{-0.5em}
\section{Conclusion}\label{sec:conclusion}
\vspace{-0.5em}
In this paper, we introduced five ordering strategies for the process discovery algorithm using synthesis rules~\cite{Huang2022Pads_1st}.
We investigated the impact of ordering strategies on model quality and computation time.
The results show that compared to the ordering strategy solely based on the frequency of activities, the proposed ordering strategies considered the connection between activities (Breadth-First-Search-based and Depth-First-Search-based) have superior performance w.r.t. time and model quality respectively.
It is shown in the result that the introduced BFS-based ordering strategies can speed up the computation.
Nevertheless, the overall discovery time of the Synthesis Miner is still not comparable to the state of the art despite being able to discover models with better quality.
Therefore, for future work, we plan to speed up the Synthesis Miner by further exploiting the log heuristics and investigating more sophisticated ordering strategies.
Another direction for improvement is the ability to cope with infrequent behaviors as we use the most frequent trace variants to capture the mainstream process.
It would be valuable to introduce a filtering mechanism to the Synthesis Miner so that it can directly work on the original log without depending on pre-filtering the log.

\subsubsection{Acknowledgements.}
\vspace{-0.5em}
We thank the Alexander von Humboldt (AvH) Stiftung for
supporting our research.
\vspace{-0.5em}
\bibliographystyle{splncs04}
\bibliography{myrefs}
%
%
%




\end{document}